\documentclass[preprint,12pt]{elsarticle}
\pdfoutput=1

\usepackage{graphicx}
\usepackage[colorlinks=true, citecolor=blue, linkcolor=blue]{hyperref}
\usepackage{array}
\graphicspath{{}}
\DeclareGraphicsExtensions{.pdf,.png}

\usepackage{amssymb}

\usepackage{amsmath}
\usepackage[textsize=tiny]{todonotes}
\usepackage{url}

\journal{arXiv}

\begin{document}

\begin{frontmatter}

\title{Contact-line pinning controls how quickly colloidal particles equilibrate with liquid interfaces}

\author[label1]{Anna Wang}
\author[label2,label3]{Ryan McGorty}
\author[label2,label4]{David M. Kaz}
\author[label1,label2]{Vinothan N.~Manoharan\corref{cor1}}
\ead{vnm@seas.harvard.edu}

\address[label1]{School of Engineering \& Applied Sciences, Harvard University, 29 Oxford Street, Cambridge, MA 02138, USA}
\address[label2]{Department of Physics, Harvard University, 17 Oxford Street, Cambridge, MA 02138, USA}
\address[label3]{Present address: Department of Physics and Biophysics, University of San Diego, San Diego, CA 92110 USA}
\address[label4]{Present address: Agilent Technologies, Santa Clara, CA 95051 USA}
\cortext[cor1]{Corresponding author}

\begin{abstract}
Previous
  experiments have shown that spherical colloidal particles relax to
  equilibrium slowly after they adsorb to a liquid-liquid interface,
  despite the large interfacial energy gradient driving the
  adsorption. The slow relaxation has been explained in terms of
  transient pinning and depinning of the contact line on the surface
  of the particles.  However, the nature of the pinning sites has not
  been investigated in detail.  We use digital holographic microscopy
  to track a variety of colloidal spheres---inorganic
  and organic, charge-stabilized and sterically stabilized, aqueous
  and non-aqueous---as they breach liquid interfaces. We find that
  nearly all of these particles relax logarithmically in
  time over timescales much larger than those expected from viscous
  dissipation alone.  By comparing our results to theoretical models
  of the pinning dynamics, we infer the area per defect to be on the
  order of a few square nanometers for each of the colloids we
  examine, whereas the energy per defect can vary from a few $kT$ for
  non-aqueous and inorganic spheres to tens of $kT$ for aqueous
  polymer particles.  The results suggest that the likely pinning
  sites are topographical features inherent to colloidal
  particles---surface roughness in the case of silica particles and
  grafted polymer ``hairs'' in the case of polymer particles.  We
  conclude that the slow relaxation must be taken into account in
  experiments and applications, such as Pickering emulsions, that
  involve colloids attaching to interfaces. The effect is particularly
  important for aqueous polymer particles, which pin the contact line
  strongly.
\end{abstract}


\end{frontmatter}

\section{Introduction}
The strong binding of colloidal particles to interfaces is exploited in
a range of applications. Particles can stabilize oil-water interfaces in
Pickering emulsions,\cite{pickering_spencer_umfreville_emulsions_1907}
which are used in food,\cite{timgren_starch_2011,
  destribats_emulsions_2014} oil recovery,\cite{zhang_nanoparticle_2010}
pharmaceuticals, and cosmetics.\cite{marku_characterization_2012}
Oil-water interfaces can also be used to scaffold the assembly of
particles into colloidosomes,\cite{dinsmore_a.d._colloidosomes:_2002}
Janus particles,\cite{paunov_novel_2003}
monolayers,\cite{retsch_nanofab_2009} and photolithography
masks.\cite{isa_particle_2010} Because the driving force for adsorption
is large---the adsorption of a single particle reduces the interfacial
energy of the system by many times the thermal energy $kT$---it is
sometimes (and often tacitly) assumed that such particles reach their
equilibrium contact angle rapidly once they breach the interface.
Indeed, if viscous drag were the only force opposing the interfacial
energy gradient, particles would relax to equilibrium exponentially with
a time constant on the order of a
microsecond.\cite{colosqui_colloidal_2013}

However, when Kaz, McGorty, and coworkers~\cite{kaz_physical_2012}
directly measured the adsorption dynamics of polystyrene microspheres
at an interface between water/glycerol and oil, they found that the
particles relaxed toward equilibrium logarithmically, not exponentially.
Furthermore, the relaxation was so slow that the time projected for the
particles to reach the equilibrium contact angle of 110$^\circ$ was
months to years---far longer than typical experimental timescales.
Later, Coertjens and coworkers~\cite{coertjens_contact_2014} directly
imaged polymer particles at vitrified interfaces and found that the
average contact angle increased an hour after adsorption. Kaz \textit{et
  al.\@} proposed that the slow relaxation is due to pinning and
unpinning of the contact line on nanoscale heterogeneties (``defects'')
on the particle surfaces. The pinning and unpinning events contribute to
a larger dissipation of energy than viscosity alone. Using a model of
contact-line hopping based on molecular-kinetic theory (described in the
Background section below), they were able to infer the sizes of the
defects.

More recent work has elucidated and expanded on how contact line pinning
affects the dynamics of particles at interfaces. Colosqui and
coworkers~\cite{colosqui_colloidal_2013} developed a model based on
Kramer's theory for the full equilibrium dynamics of the particles,
including not only the logarithmic regime, but also the dynamics shortly
after the breach and close to equilibrium. As we describe below in the
Background section, this model can be fit to experimental data to
estimate the pinning energy per defect. Other work examines the effect
of pinning on particle dynamics lateral to an interface. Recent
experimental studies by Boniello \textit{et
  al.}~\cite{boniello_brownian_2015} indicate that the lateral diffusion
of colloidal particles at a fluid interface is likely slowed by
transient pinning events. Sharifi-Mood and
coworkers~\cite{sharifi-mood_curvature_2015} showed that strong pinning
can locally distort the interface around a colloidal particle, affecting
how particles migrate on a curved surface.

These studies highlight the importance of contact-line pinning for
understanding the dynamics of colloids at interfaces. The observed slow
relaxation has direct consequences for the applications we list above:
in a collection of identical particles at an interface, such as the
surface of a Pickering emulsion droplet, the particles can have
different contact angles that change over time. Because the contact
angle of a particle determines the length of the three-phase contact
line and how much of the particle is exposed to the aqueous or oil
phases, it affects the capillary~\cite{kralchevsky_particles_2001} and
electrostatic interactions between
particles.\cite{mcgorty_colloidal_2010} Contact angles that change over
time might help explain the heterogeneous
pair-interactions~\cite{park_heterogeneity_2010, park_direct_2008} and
long-ranged attractions observed between identically charged
particles.\cite{nikolaides_m.g._electric-field-induced_2002} For the
particular case of Pickering emulsions, the emulsion type (water-in-oil,
or oil-in-water) also depends on the contact
angle,\cite{binks_direct_2013} and so a changing contact angle might
change the emulsion type and stability over time.

Here we focus on understanding how ubiquitous the pinning is and what
causes it. To do this, we follow the approach of Kaz \textit{et
  al.}~\cite{kaz_physical_2012} and Wang \textit{et
  al.}~\cite{wang_relaxation_2013} and use digital holographic
microscopy, a fast three-dimensional imaging technique, to measure the
motion of spherical particles as they breach liquid interfaces. However,
here we examine a much wider variety of particles and surface
functionalities. We find that charge-stabilized polymer spheres
(including a variety of emulsion-polymerized particles),
surfactant-stabilized polymer spheres, and large (several micrometers in
diameter) silica spheres all relax logarithmically to equilibrium,
though some systems, including oil-dispersed PMMA particles and smaller
silica spheres, reach equilibrium on experimental timescales. By fitting
models to the data, we are able to extract details about the pinning
sites. For example, we find that the heterogeneities on
aqueous-dispersed polymer particles pin the contact line with an order
of magnitude more energy than those on other particles, resulting in a
longer logarithmic regime. We conclude that the likely pinning sites are
nanoscale topographical features such as polymer ``hairs.''

\begin{figure*}
\centering
  \includegraphics{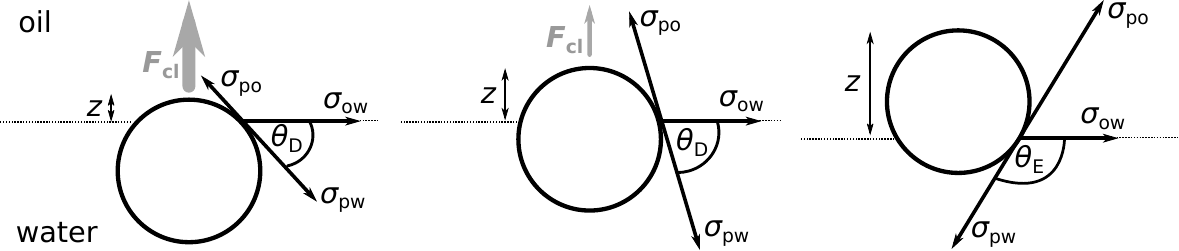}
  \caption[Interfacial forces on a particle at an interface.]{When the
    particle first breaches the interface, the unbalanced interfacial
    tensions cause the particle to move. This unbalanced force
    decreases as the particle approaches equilibrium, where the
    dynamic contact angle $\theta_\mathrm{D}$ reaches its equilibrium
    value $\theta_\mathrm{E}$ and the force $F_\mathrm{cl}$ goes to
    zero.}
  \label{fcl}
\end{figure*}

\section{Background}
In this section we describe the theories that have been developed to
explain the slow relaxation of colloidal particles at interfaces, and
how fitting them to experimental data reveals details of the pinning
dynamics. The logarithmic trajectories observed by Kaz \textit{et
  al.}~\cite{kaz_physical_2012} can be explained by using molecular
kinetic theory (MKT) to model the motion of the contact line as a
dynamic wetting process.\cite{blake_kinetics_1969} In
  this model, as the contact line moves across the surface of the
  particle, it encounters defects of area $A$ that each pin it with
  energy $\Delta U$. The contact line requires a thermal ``kick'' to
  keep it moving toward equilibrium: once it unpins from one defect, it
  can then move along the particle until it gets caught on another
  defect. The characteristic length the contact line traverses before
  reaching another defect is $\ell$ =
  $A$/$p$,~\cite{colosqui_colloidal_2013} where $p$ is the perimeter of
  the contact line. This model explains why the particle motion appears
  continuous in the experiments: in practice, $\ell$ is on order of
  picometers~\cite{colosqui_colloidal_2013}, much smaller than
  displacements that we can measure. The model also explains why the
  particle slows as it progresses through the interface: the driving
  force decreases as the particle gets closer to equilibrium, while the
  pinning energy and the density of defects remain the same
  (Figure~\ref{fcl}).

The activated hopping of the contact line results in much more
dissipation than that predicted from hydrodynamics. If hydrodynamics
were the only relevant effect, we would expect the particle to follow an
exponential path to equilibrium with a timescale $T_\mathrm{D} \approx
\eta r / \sigma_\mathrm{ow}$, where $\eta$ is a weighted average of the
viscosities of the two fluids, $r$ the radius of the particle, and
$\sigma_\mathrm{ow}$ the interfacial tension between oil and
water.\cite{colosqui_colloidal_2013} For a 1-$\mu$m-radius particle at a
water-alkane interface, $T_\mathrm{D}$ is approximately 0.1 $\mu$s,
which is several orders of magnitude smaller than the times observed in
experiments.\cite{kaz_physical_2012}

A model based on MKT, presented in the supplementary information section
of Kaz \textit{et al.},\cite{kaz_physical_2012} captured the
experimentally observed dynamics from 10$^{-2}$--10$^2$ s after the
breach---the point where the particle first comes into contact with the
interface and a three-phase contact line is formed. This model is not
valid for shorter times, where the length of the contact line rapidly
increases; instead, it is intended to model the behavior in the
logarithmic regime, where the contact line perimeter changes slowly with
time. By fitting the model to the data, the authors inferred that the
area per pinning defect was on the order of a few square nanometers.
This value is larger than the molecular scales the theory was derived
for, but there are other successful applications of MKT to surfaces with
defects larger than 1 nm$^2$.\cite{rolley_dynamics_2007,
  snoeijer_moving_2013}

To show how the area per defect affects the dynamics, we present a brief
derivation of the model from Kaz \textit{et al.}\cite{kaz_physical_2012}
We model the activated hopping process using an Arrhenius equation for
the velocity of the contact line.\cite{blake_kinetics_1969} Far from
equilibrium, we can neglect backward hops. In this case, the velocity of
the contact line tangent to the particle is given by
\begin{equation}
V = V_0 \exp\left(-\frac{\Delta U}{kT}+\frac{F_\mathrm{cl}(t)A}{2kT}\right)
\label{eqn:v}
\end{equation}
where $V_0$ is a molecular velocity scale, and $kT$ is the thermal
energy. The force per unit length on the contact line, $F_\mathrm{cl}$,
is determined by the tangential component of the oil-water
($\sigma_\mathrm{ow}$), particle-oil ($\sigma_\mathrm{po}$), and
particle-water ($\sigma_\mathrm{pw}$) interfacial tensions
(Figure~\ref{fcl}):
\begin{equation}
\begin{split}
  F_\mathrm{cl} &= {\sigma}_\mathrm{ow}\cos{\theta}_\mathrm{D}(t)+\sigma_\mathrm{pw}-\sigma_\mathrm{po}  \\
  &= {\sigma}_\mathrm{ow}\left(\cos{\theta}_\mathrm{D}(t)-\cos{\theta}_\mathrm{E}\right)
\end{split}
\label{eqn:f}
\end{equation}
where $\theta_\mathrm{D}$ is the dynamic contact angle. 

Substituting Equation~\eqref{eqn:f} into Equation~\eqref{eqn:v} and
rewriting the resulting equation of motion in terms of the observable
axial coordinate $z$, we obtain
\begin{equation}
\dot{z} = \nu r \sin(\theta_\mathrm{D}) \exp\left(\frac{A\sigma_\mathrm{ow}z}{2rkT}\right)
\newline = \nu \sqrt{z(2r-z)} \exp\left(\frac{A\sigma_\mathrm{ow}z}{2rkT}\right)
  \label{eqn:zdot}
\end{equation}
where $r$ is the radius of the particle and
\begin{equation*}
  \nu =
  (V_0/r)\exp\left(-\Delta U/kT+\left(1-\cos\theta_\mathrm{E}\right)A\sigma_\mathrm{ow}/2kT\right).
\end{equation*}
In deriving Equation~\eqref{eqn:zdot} we have assumed that the interface
remains flat at all times. We have also let $z=0$ when the particle
first touches the interface (at $\theta_\textnormal{D}$ = 0), from which
we obtain $z = r\left(1-\cos\theta_\textnormal{D}\right)$ and $V =
r\dot{\theta}_\textnormal{D} = \dot{z}/\sin\theta_\textnormal{D}$.

When the particle is close to equilibrium, we can expand around the
equilibrium contact angle and solve the resulting differential equation
to obtain the equation of motion
\begin{equation*}
z \approx \frac{2rkT}{A\sigma_\mathrm{ow}} \log \left( \frac{A\sigma_\mathrm{ow}}{2rkT}\nu r \left(\sin\theta_\mathrm{E}\right) t \right)
\end{equation*}
which we can rewrite as
\begin{equation}
 \label{eqnsimple}
\frac{z}{r} \approx \frac{2kT}{A \sigma_\mathrm{ow}} \log \frac{t}{t_0} + C \ ; \quad C = \frac{2kT}{A \sigma_\mathrm{ow}} \log \left(\frac{A\sigma_\mathrm{ow}}{2rkT} \nu r \left(\sin\theta_\mathrm{E}\right) t_0 \right).
\end{equation}
Equation~\eqref{eqnsimple} shows that the trajectory of the particle is
approximately logarithmic in time. We can infer the area per defect $A$
from the slope of a plot of $z$ as a function of $\log t$. We cannot
determine the constant $C$---and hence the pinning energy per defect
$\Delta U$, which is embedded in $\nu$---by fitting this model to the
data. We therefore choose an arbitrary $t_0$ ($t_0$=1 s).

To determine the pinning energy per defect, $\Delta U$, we must observe
where the logarithmic regime begins. Colosqui \textit{et
  al.},\cite{colosqui_colloidal_2013} using Kramer's
theory,\cite{kramers_brownian_1940} showed that particles having
heterogeneous surface defects initially relax exponentially and then
logarithmically. The models from Kaz \textit{et al.} and Colosqui
\textit{et al.} are mathematically
equivalent~\cite{colosqui_colloidal_2013} when the dynamic contact angle
$\theta_\mathrm{D}$ is approximately $\pi$/2. The area per defect $A$
from Kaz \textit{et al.} is related to the length scale $l$ from
Colosqui \textit{et al.} by $A \sim 2{\pi}R^*l$, where $R^*$ is the
radius of the contact line when the particle is at $z_\mathrm{C}$, and
$z_\mathrm{C}$ is the height at which the relaxation changes from
exponential to logarithmic. The crossover point between exponential and
logarithmic regimes can be used to infer $\Delta U$, if the equilibrium
height of the particle $z_\mathrm{E}$ is known or can be estimated:
\begin{equation}
z_\mathrm{E}-z_\mathrm{C} = \frac{\Delta U \pi R^*}{2\sigma_\mathrm{ow}A}.
 \label{crossover}
\end{equation}

To analyze our experimental data we fit Equation~\eqref{eqnsimple} to
the logarithmic regime to obtain $A$ and then use
Equation~\eqref{crossover} to determine the defect energy $\Delta U$. We
note that these models capture only the gross features of the
trajectories. A more recent model~\cite{rahmani_colloidal_2016} expands
on the model of Colosqui \textit{et al.} to include extra dissipative
effects. This model captures both the short- and long-time behavior of
the experimental results from Kaz \textit{et al.} well. Here, because we
are interested primarily in the two parameters $A$ and $\Delta U$, we do
not seek to capture the full time-dependence of the adsorption process,
and we examine our results in the context of the simpler models from Kaz
\textit{et al.} and Colosqui \textit{et al.}.

\section{Materials and methods}

\subsection{Particles and interfaces}
To determine what kinds of surface features affect how a particle
relaxes to equilibrium, we track particles with a variety of different
surface properties as they breach an interface between an aqueous phase
and oil. The types of particles we examine are listed in
Table~\ref{table:polymer}. They include 1.9-$\mu$m-diameter
charge-stabilized sulfate- and carboxyl-functionalized polystyrene (PS,
Invitrogen), 2.48-$\mu$m-diameter sulfate-functionalized poly(methyl
methacrylate) (PMMA, Bangs Laboratories, synthesized by emulsion
polymerization), 1.7-$\mu$m-diameter poly\-vinyl\-alcohol-stabilized PS
(synthesized according to the procedure in Paine \textit{et
  al.}~\cite{paine_dispersion_1990}), 3.7-$\mu$m-diameter
poly\-vinyl\-pyrrolidone-stabil\-ized PMMA (synthesized according to the
procedure in Cao \textit{et al.}~\cite{cao_micron_2000}), and
1.0-$\mu$m-diameter bare silica microspheres with SiOH surface groups
(Bangs Laboratories). We centrifuge and wash each suspension ten times
in deionized water (EMD Millipore, resistivity = 18.2 M$\Omega \cdot$cm)
to remove contaminants and surface-active compounds, then dilute them
for use in experiments.

We also examine several different types of oil-dispersible particles:
1.0-$\mu$m-diameter (Bangs Laboratories) and 4.0-$\mu$m-diameter
(AngstromSphere) silica microspheres, both with SiOH surface groups,
1.1-$\mu$m-diameter poly\-dimethyl\-siloxane-stabil\-ized PMMA
(synthesized according to the procedure in Klein \textit{et
  al.}~\cite{klein_preparation_2003}), and 1.6-$\mu$m-diameter
poly(12-hydroxystearic acid)-stabil\-ized PMMA particles (synthesized
according to the procedure in Elsesser \textit{et
  al.}~\cite{elsesser_revisiting_2010}). We wash the particles five
times in decane ($\geq$99\%, anhydrous, Sigma-Aldrich) to remove
possible contaminants. We discard any macroscopic colloidal aggregates
and keep the freely suspended particles for experiments.

\begin{table}[h]
\centering
\small
\caption{Particles used in breaching experiments, along with the
  shortened names we use to refer to them in the text.}
\begin{tabular}{ p{2.5cm}  p{5cm} p{2.5cm} m{1cm}}
  Name & Particle & Phase & Diameter ($\mu$m) \\  \hline
  PMMA & Sulfate-functionalized PMMA & aqueous & 2.48 \\
  PVP-PMMA & Polyvinylpyrrolidone-stabilized PMMA & aqueous & 3.67 \\
  sulfate-PS & Sulfate-functionalized PS & aqueous & 1.88 \\
  carboxyl-PS & Carboxylate-functionalized PS & aqueous & 1.88 \\
  PVA-PS & Polyvinylalcohol-stabilized PS & aqueous & 1.65 \\
  PDMS-PMMA & Polydimethylsiloxane-stabilized PMMA & oil & 1.1 \\
  PHSA-PMMA & Poly(12-hydroxystearic acid)-stabilized PMMA & oil & 1.6 \\
  silica & Bare silica  & aqueous or oil & 1.0 \\
  large silica & Bare silica & oil & 4.0 \\
\end{tabular}
\label{table:polymer}
\end{table}

We prepare different aqueous phases from deionized water, anhydrous
glycerol ($\geq$99\%, Sigma-Aldrich), pure ethanol (100\%, KOPTEC), and
hydrochloric acid (Fluka). All of the aqueous solutions contain 100 mM
NaCl (99.5\%, EMD) to screen any electrostatic repulsion between the
particle and the interface~\cite{kaz_physical_2012}. For the oil phase
we use decane ($\geq$99\%, anhydrous, Sigma-Aldrich) which is first
filtered through a PTFE membrane filter (Acrodisc). The different
liquid-liquid interfaces we use in experiments are summarized in
Table~\ref{table:liquids}.

\begin{table}
\centering
\small
\caption{Aqueous phase-decane interfaces used, along with the
  name we use to refer to them.}
\begin{tabular}{l p{3cm} l}                  
  Name & Aqueous phase (index) & Oil (index) \\  \hline
  water/glycerol & 59\% w/w glycerol in water (1.411) & decane (1.411)  \\
  water & Water (1.333) & decane (1.411) \\
  water/ethanol & 10\% v/v ethanol in water (1.380) & decane (1.411) \\
\end{tabular}
\label{table:liquids}
\end{table}

We measure the interfacial tension between water/glycerol and decane
using the pendant drop method.\cite{rotenberg_determination_1983,
  touhami_modified_1996} A 1 mL syringe (Sigma-Aldrich) with a blunt end
syringe needle (18 gauge, Kimble) is filled with the aqueous phase. The
needle is then submerged into a disposable cuvette (VWR) filled with
decane. A droplet of the aqueous phase is slowly injected into decane
while images are recorded. The profile of the droplet is analyzed from
the images to determine the interfacial tension.

\subsection{Sample preparation}

Our custom-made polyether ether ketone (PEEK) sample cells are glued to
a glass coverslip with UV-cured epoxy (Norland 60). A detailed
description of their fabrication can be found in the supplemental
information of Kaz \textit{et al.}\cite{kaz_physical_2012} Using these
cells, we create a stable oil-water interface consisting of a
30--80-$\mu$m-thick aqueous phase and a 2--3-mm-thick decane superphase.
We use No.~1 coverslips (VWR) so that the interface is within the
working distance of an oil-immersion objective (NA=1.4, Nikon CFI Plan
Apo VC 100$\times$) or water-immersion objective (NA=1.2, Nikon CFI Plan
Apo VC 60$\times$). We bake all glassware used to handle the colloidal
particles and fluids in a pyrolysis oven (Pyro-Clean Tempyrox) to
incinerate organics, then sonicate and wash the glassware with deionized
water. This protocol is designed to eliminate interfacially-active
contaminants.

We place the sample cell on a Nikon TE-2000 inverted microscope. We
focus 5--15 $\mu$m below the interface to capture holograms of
individual particles as they breach. If we start with particles that are
suspended in the aqueous phase, we push them toward the interface using
radiation pressure (force less than 1 pN) from out-of-focus optical
tweezers, as shown in Figure \ref{dhm}. If the particles are suspended
in oil, we simply allow them to sediment toward the interface.

\subsection{Tracking particles with Digital Holographic Microscopy}

\begin{figure}
\centering
  \includegraphics{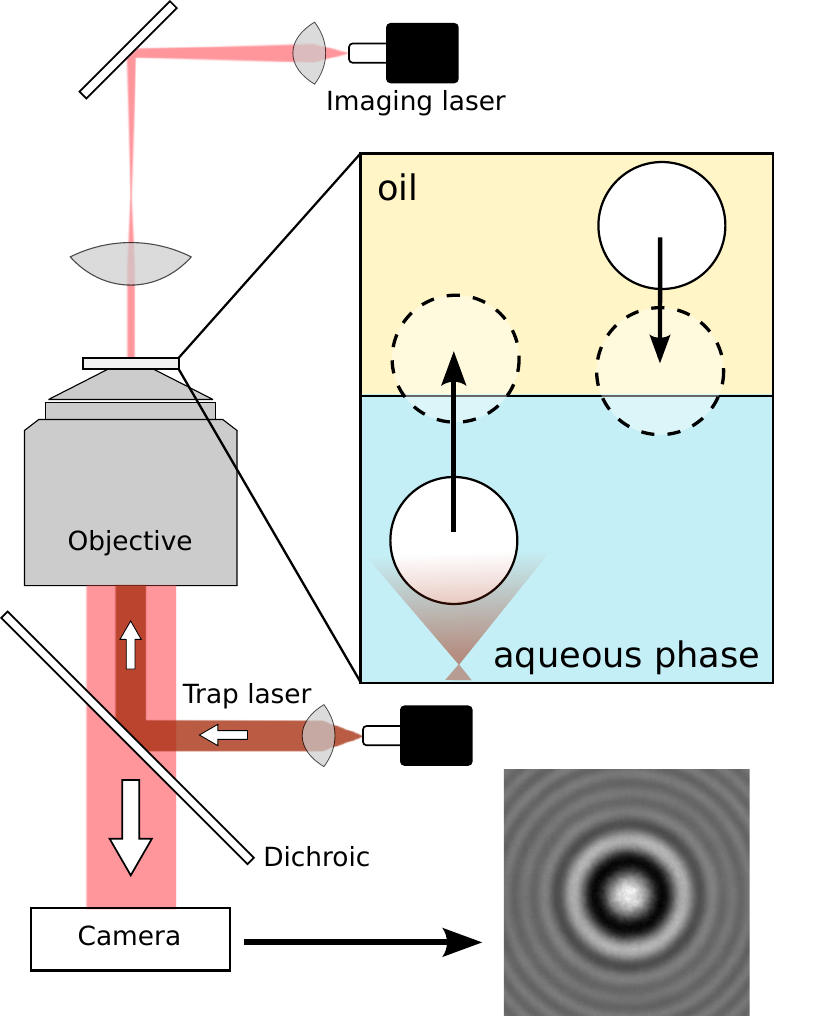}
  \caption[Experimental apparatus.]{Experimental setup. The sample sits
    on an inverted microscope and is illuminated from above with a
    collimated 660 nm laser. The hologram formed by the interference of
    the scattered light from the sample with the undiffracted beam is
    then captured on a camera. We push spheres from the aqueous phase
    toward the interface with an 830 nm laser. To observe the breaching
    and relaxation of a particle from the aqueous phase, we push it
    gently toward the interface using optical tweezers, and measure its
    trajectory using holographic microscopy. To observe particles
    breaching from the oil phase, we simply let the particles fall to
    the interface.}
  \label{dhm}
\end{figure}

We use an in-line digital holographic microscope, based on a modified
Nikon TE-2000 inverted microscope, to track the particles with high
temporal and spatial resolution in all three dimensions
(Figure~\ref{dhm}). We illuminate samples with a 660 nm imaging laser
(Opnext HL6545MG) that is spatially filtered through a single-mode
optical fiber (OzOptics SMJ-3U3U-633-4/125-3-5) and collimated. We use a
counterpropagating 830 nm trap laser (Sanyo DL8142-201), which is
spatially filtered through a single-mode optical fiber (OzOptics
SMJ-3U3U-780-5/125-3-5), to push particles toward the interface.

The imaging beam (typically 50 mW power) scatters from the sample and
interferes with undiffracted light to produce an interference pattern,
or hologram. After passing through the objective, holograms are recorded
on a monochrome CMOS camera (Photon Focus MVD-1024E-160-CL-12), captured
with a frame grabber (EPIX PIXCI E4), and then saved to disk for further
processing. We use a short camera exposure time, 20 $\mu$s, to minimize
motion blur, and we capture holograms at up to 2000 frames per second,
giving us sub-millisecond time resolution. A background image, taken in
a part of the sample with no particles, is also recorded and divided
from each time-series of holograms to remove artifacts arising from
scattering from imperfections on the camera, lenses, and mirrors.

The background-divided holograms are analyzed using our open-source
software package HoloPy
(\href{http://manoharan.seas.harvard.edu/holopy}{http://manoharan.seas.harvard.edu/holopy}).
To extract the particle trajectories, we fit the Lorenz-Mie scattering
model to each background-divided hologram, as described in Fung
\textit{et al.}, \cite{fung_jqsrt_2012} following the work of Ovryn and
Izen~\cite{ovryn_imaging_2000} and of Lee and
coworkers.\cite{lee_characterizing_2007}

The accuracy of the Lorenz-Mie model depends on the refractive index
mismatch between the two liquid phases. The Lorenz-Mie scattering
solution used to analyze the holograms is exact only for particles in an
optically homogeneous medium. Exact light scattering solutions for
particles straddling an optically discontinuous boundary do not exist.
Therefore, to determine the position of bound particles with maximum
accuracy, we index-match the aqueous phase to decane ($n$ = 1.41) by
mixing anhydrous glycerol ($\geq$99\%, Sigma-Aldrich) with water to make
a solution of 59\% w/w glycerol so that the system is optically
continuous. This index-matching also eliminates reflections from the
fluid-fluid interface, which would produce additional interference.

For some of the experiments, we cannot index match the aqueous medium to
the oil phase. Because silica particles typically have a refractive
index of 1.42 at our imaging wavelength (660 nm) and a high density
compared to water, we cannot obtain sufficient radiation pressure to
push them toward the interface if they are submerged in an aqueous
medium with $n$ = 1.41. Instead, we disperse them in water ($n$ = 1.33)
so that the refractive index contrast between the particles and medium
is large enough for us to manipulate them with the trapping laser. In
our analysis, we allow the refractive index of the particle relative to
that of the medium to vary during the fit, which helps compensate for
the change in medium index as the particle moves through the interface.
In this way we are able to measure the approximate relaxation behavior
of the silica spheres.

Because microscope objectives and their immersion fluids are designed to
image objects in two dimensions, a difference in refractive index
between the immersion oil and the medium leads to spherical aberration,
which distorts distances in the axial direction~\cite{egner_2006} and
compromises the positioning accuracy. To mitigate this effect, we use an
immersion oil with $n$=1.4140 (Series AA, Cargille) with our 100$\times$
oil-immersion objective for samples where we index-match the aqueous
phase to decane ($n$=1.41). In experiments where pure water ($n$=1.33)
is the aqueous phase, we use a water-immersion objective with water as
the immersion fluid.

\section{Results}

\subsection{Slow relaxation is not particular to a water/glycerol-decane interface}
We begin by showing that the slow relaxation of particles at an
interface is not particular to the decane-water/glycerol system of Kaz
\textit{et al.}\cite{kaz_physical_2012} In that work, the aqueous phase
was designed to match the refractive index of decane yet retain an
interfacial tension and Debye screening length similar to water. Here we
track sulfate-PS particles as they approach interfaces from different
aqueous solutions (Table \ref{table:liquids}).

\begin{figure}
\centering
  \includegraphics{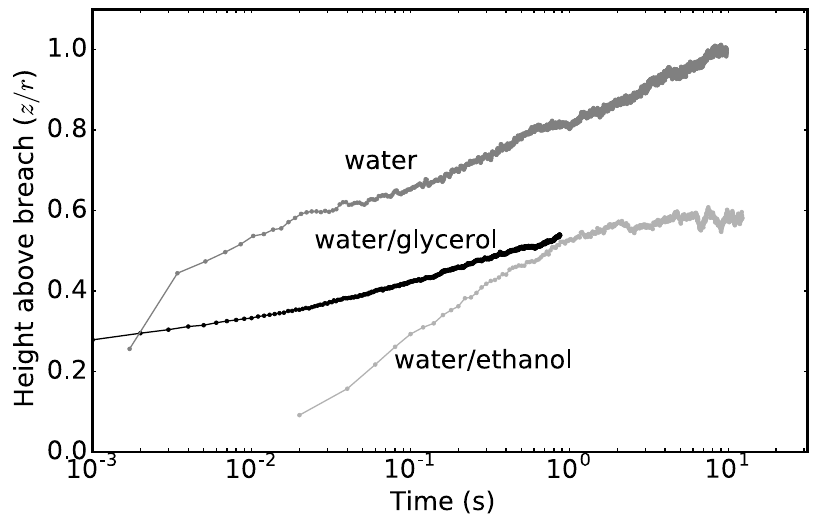}
  \caption[Typical trajectories of different particles as they breach
  various aqueous-oil interfaces.]{Typical trajectories of sulfate-PS
    particles as they breach various aqueous phase-oil interfaces, as
    listed in Table \ref{table:liquids}. The distance between the top
    surface of the particle and the interface is shown as a function of
    the time after the breach. All of the trajectories show logarithmic
    relaxation. We define $z$ = 0 $\mu$m as the height at which the
    particle and interface first touch.}
  \label{nonmatched_log}
\end{figure}

The motion of the polystyrene particles through the interface is
approximately logarithmic with time in all of the systems, as shown in
Figure~\ref{nonmatched_log}. We note that the different starting times
for the plots are an artifact of the logarithmic time-axis and the
different frame rates used to acquire the data. The differences in
slopes for the trajectories are due in part to the different
refractive-index mismatches (and hence tracking errors) in the three
systems, and in part to the different interfacial tensions and
dielectric constants in the systems. However, we do not expect any of
these effects to change the functional (logarithmic) relationship
between height and time. Therefore, we conclude that the slow dynamics
are not unique to the water/glycerol and decane system studied in Kaz
\textit{et al.}~\cite{kaz_physical_2012} and are potentially relevant to
a variety of other liquids.

\subsection{Topographical features on polymer particles pin the
  contact line}

Though colloidal particles may appear smooth under optical and even
scanning electron microscopy, the particle surfaces contain nanoscale
heterogeneities such as charges, asperities and, in the case of polymer
particles, polymer ``hairs.''\cite{rosen_saville_1990,
  rosen_heatparticles_1992} To determine which of these features is
responsible for the slow relaxation, we return to the index-matched
water/glycerol and decane system and quantitatively measure how
particles with different surface features breach the interface. From the
trajectories we determine the area per defect, $A$, by fitting
Equation~\eqref{eqnsimple} to the logarithmic regime of the measured
trajectories. Kaz and coworkers found that $A$ was on the order of the
area per charge group for sulfate- ($A\approx$ 5 nm$^2$), carboxyl-
($A\approx$ 3 nm$^2$), amidine- ($A\approx$ 15 nm$^2$), and
carboxylate-modified-latex ($A\approx$ 25 nm$^2$) spheres. These results
suggest that the charges themselves, or some surface features associated
with the charges, could be the pinning sites.

To understand if and how the charges influence the pinning, we do
experiments on PS-carboxyl spheres suspended in a 59\% glycerol in water
solution containing 100 mM NaCl. We work with carboxyl-functionalized
spheres because the pKa is higher than that of sulfate-functionalized
spheres, so that the charge can be adjusted by changing the pH over a
moderate range. We add acid to the suspensions and measure the zeta
potentials of the particles using a Beckman Coulter DelsaNano C zeta
potentiometer. The zeta potential decreases by a factor of about four
over a range of acid concentrations from 0 to 10$^{-3}$ M
(Table~\ref{table:acidzeta}). Measurements of the interfacial tension
using the pendant drop method with a slight index mismatch, caused by
increasing the water content in the aqueous phase by about 1\% w/w,
confirm that the interfacial tension does not vary with acid
concentration.

\begin{table}
\centering
\small
\caption{Zeta potentials of PS-carboxyl latex at different acid
  concentrations.}
\begin{tabular}{c c }                  
  Concentration of HCl (M) & Zeta potential (mV) \\  \hline
  0 & -95 $\pm$ 10 \\
  10$^{-5}$ & -68 $\pm$ 2 \\
  10$^{-4}$ & -60 $\pm$ 2 \\
  10$^{-3}$ & -24 $\pm$ 1 \\
\end{tabular}
\label{table:acidzeta}
\end{table}

\begin{figure}
\centering
  \includegraphics{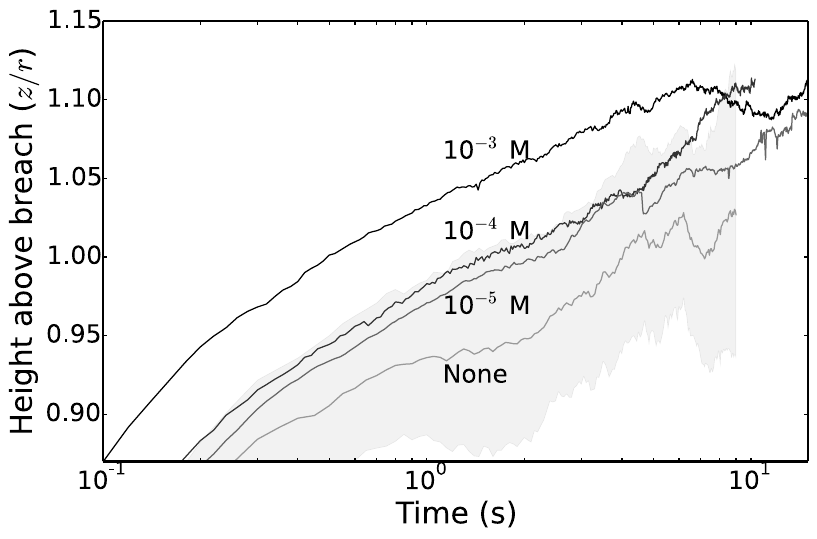}
  \caption[Trajectories of carboxyl-functionalized latex particles in
  solutions of varying acid concentration.]{Trajectories of
    carboxyl-functionalized latex particles in solutions of varying acid
    concentration (HCl concentrations are marked above each curve).
    Lines are the average of five particle trajectories at each
    concentration. The gray shaded region shows the uncertainty in the
    zero-HCl-concentration measurement, as determined by the standard
    deviation on the five trajectories. It is representative of the
    uncertainties at the other concentrations.}
  \label{acid}
\end{figure}

We find that at any given time after the particles breach, particles
submerged in higher acid concentrations are at larger heights
(Figure~\ref{acid}). There are two possible interpretations of this
observation in the context of Kramer's theory and
Equation~\eqref{crossover}: either the energy of the defect decreases
with acid concentration, or the equilibrium contact angle increases with
acid concentration. These two quantities cannot be determined
independently using either of the two dynamic
models;\cite{colosqui_colloidal_2013, kaz_physical_2012} however, it
stands to reason that a smaller surface charge should increase the
hydrophobicity of the particles and thus their equilibrium contact
angle.

To better understand the nature of the pinning sites, we fit
Equation~\eqref{eqnsimple} to the the logarithmic regime in our data
(Figure~\ref{acid}). The area per defect, which influences the slope of
the trajectory, is between 4 nm$^2$ and 6 nm$^2$ for each of the four
samples. The areas per defect measured here and in Kaz \textit{et
  al.}~\cite{kaz_physical_2012} differ by orders of magnitude from the
roughly (100 nm)$^2$ chemical haterogeneities of polystyrene particles
measured under atomic force microscopy~\cite{chen_measured_2006}. This
disparity suggests that the defects are not the chemical patches that
are seen under surface characterization. Moreover, the fact that the
areas per defect are nearly constant, despite the large variation in
zeta potential (and hence area per charge) with acid concentration,
suggests that the pinning sites are not the charges themselves but
rather topographical features associated with the charged groups.

Prompted by a question from a reviewer of this manuscript, we also
consider whether the slow relaxation might be related to swelling of the
particles and subsequent deformation, as discussed by Park \textit{et
  al.}~\cite{park_fabrication_2010} and Tanaka \textit{et
  al.}~\cite{tanaka_novel_2010}. To determine whether the particles
swell as they come into contact with decane, we measure the refractive
index of our particles throughout the whole trajectory. If the
polystyrene particles were swelling upon contact with decane, we would
expect their refractive index to decrease, since $n_\mathrm{PS}$ is
1.59, while $n_\mathrm{decane}$ is 1.41. However, we find $n = 1.593 \pm
0.001$ before the breach and $n = 1.590 \pm 0.002$ several seconds after
the breach (where the error is the standard error from fitting
individual time points in a series). We conclude that there is no
significant swelling during the breaching process.

\begin{figure}
\centering
  \includegraphics{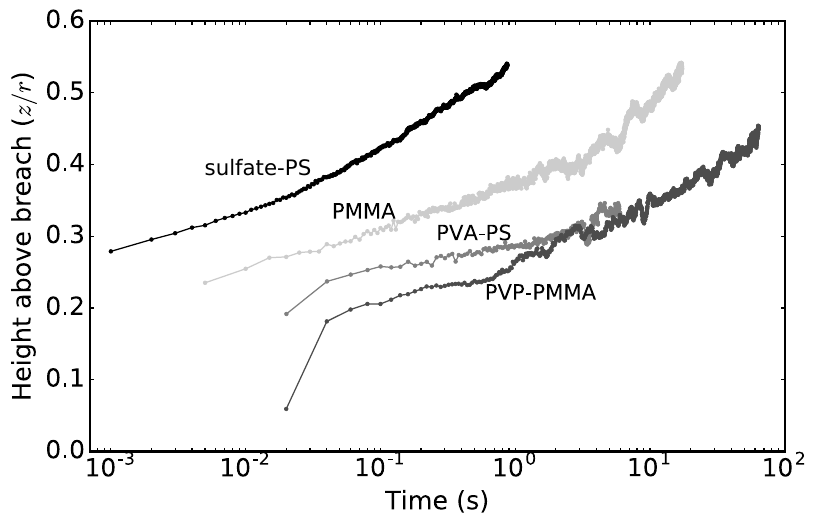}
  \caption[Typical trajectories of different polymer particles.]{Typical
    trajectories of different polymer particles as they breach
    interfaces between water/glycerol and decane. The particle details
    are listed in Table \ref{table:polymer}. The height of the particle
    above the interface as a function of the time after the breach is
    shown. All the trajectories show logarithmic relaxation.}
  \label{polymerparticles}
\end{figure}

We probe the breaching behavior of a range of other polymer particles to
gain further insights. We examine both charge-stabilized and
sterically-stabilized particles. The charge-stabilized particles include
sulfate-PS and PMMA, both of which are synthesized by emulsion
polymerization, while the sterically-stabilized particles include PVA-PS
and PVP-PMMA, both of which are synthesized by dispersion polymerization
(Table~\ref{table:polymer}).

All of these polymer particles relax logarithmically after breaching, as
shown in Figure~\ref{polymerparticles}. We fit
Equation~\eqref{eqnsimple} to the data to yield $A$ = 4.6--11 nm$^2$ for
the particles. Using Equation~\eqref{crossover}, we calculate the
pinning energies using $\sigma_\mathrm{ow}$ = 37 mN/m, $T$ = 295 K, and,
for sulfate-PS, an equilibrium contact angle of 116$^\circ \pm
10^\circ$~\cite{isa_measuring_2011, maestro_contact_2014} and
$z_\mathrm{C}/r$ = 0.3; for PVA-PS, an equilibrium contact angle of
100$^\circ \pm 20^\circ$~\footnote{Polystyrene with some PVA on the
  surface, angle taken from the measurement for ``double-cleaned''
  polystyrene in Isa \textit{et al.}~\cite{coertjens_contact_2014}} and
$z_\mathrm{C}/r$ = 0.2; and, for both types of water-dispersed PMMA
particles, an equilibrium contact angle of 90$^\circ \pm
30^\circ$~\footnote{No measurements for the equilibrium contact angle of
  aqueous-dispersible PMMA particles could be found. The PMMA particles
  from Bangs Laboratories, Inc. are expected to be more hydrophilic than
  typical polystyrene particles.} and $z_\mathrm{C}/r$ = 0.2. We find
that $\Delta U$ = 50--100 $kT$, as shown in
Table~\ref{table:polymerfits}.

\begin{table}[h]
\centering
\small
\caption[Fitted $A$ and $\Delta U$ for various polymer particles]{Fitted
  $A$ and $\Delta U$ for various polymer particles. The uncertainties in
  $\Delta U$ account for uncertainties in the values of $z_\mathrm{C}$
  and $z_\mathrm{E}$.}
\begin{tabular}{c c c}                  
  Particle type & $A$ (nm$^2$) & $\Delta U$ ($kT$)\\  \hline
  PMMA  & 7.1 & 55 $\pm$ 35 \\
  PVP-PMMA & 6.3 & 50 $\pm$ 30 \\
  sulfate-PS & 4.6 & 55 $\pm$ 5 \\
  PVA-PS & 11 & 100 $\pm$ 30\\
\end{tabular}
\label{table:polymerfits}
\end{table}

\begin{figure*}
\centering
  \includegraphics[width=\textwidth]{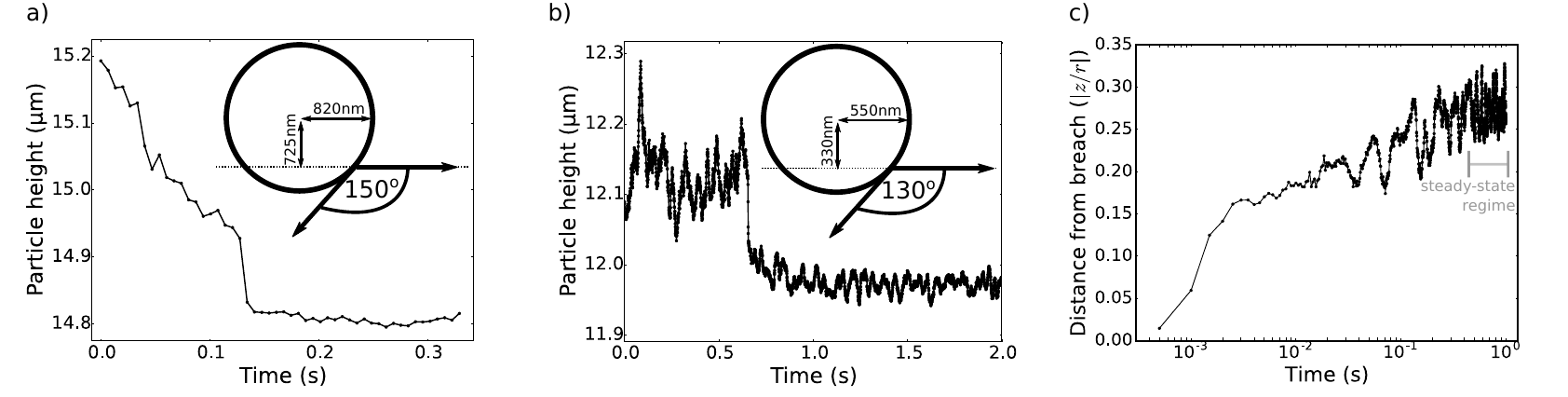}
  \caption[Sterically-stabilized oil-dispersed PMMA particles reach a
  steady-state contact angle within 1 s.]{a) A PHSA-PMMA sphere reaches
    a steady-state contact angle of 150$^\circ$. b) Fluctuations of a
    1.1-$\mu$m-diameter PDMS-PMMA particle decrease after the particle
    breaches the interface. The sphere reaches a steady-state contact
    angle of 130$^\circ$. c) The same data from b) plotted on
    semilogarithmic axes, showing the initial logarithmic relaxation
    followed by the transition to a steady-state height.}
  \label{PMMAzoom}
\end{figure*}

We also examine the relaxation of PDMS-PMMA and PHSA-PMMA particles,
both of which are sterically stabilized and dispersed in the oil phase.
We can determine the contact angle of the particles (see
Figure~\ref{PMMAzoom}) from the heights before and after they breach the
interface. Both PHSA- and PDMS-stabilized particles reach a steady-state
contact angle of 130--150$^\circ$ within a second of breaching. We find
the steady-state contact angle is 135$^\circ \pm 10^\circ$ for the
PDMS-stabilized particles, and 150$^\circ \pm 5^\circ$ for the
PHSA-stabilized particles, where the uncertainty is determined from the
standard error in the measurement of the height for five different
particles. These contact angles are close to the those measured for PMMA
particles at a water-decane interface using the freeze-fracture
shadow-casting cryoSEM technique (130$^\circ \pm 12^\circ$) and using
the gel-trapping technique (157$^\circ \pm 6^\circ$)
\cite{leunissen_electrostatics_2007, isa_measuring_2011,
  maestro_contact_2014}

The relaxation of the the sterically stabilized PMMA particles is much
faster than that of polymer spheres dispersed in the aqueous phase. To
understand this difference, we use the two dynamical models to infer the
area and pinning energy per defect. Fitting Equation~\eqref{eqnsimple}
to the first second of the breaching trajectory for the PDMS-PMMA
particle (Figure~\ref{PMMAzoom}c) yields $A$ = 8 nm$^2$. From
Equation~\eqref{crossover}, we calculate the pinning energy using
$z_\mathrm{C}/r$ = 0.16, $\sigma_\mathrm{ow}$ = 37 mN/m, $T$ = 295 K and
$R^*$ = 330 nm. We find $\Delta U$ = 4 $kT$. Thus the area per defect is
comparable to that of the aqueous-dispersed particles, but the energy
per defect is an order of magnitude smaller.

Because these particles likely have few charges, the area per defect is
too small to be comparable to the area per charged group. So in this
case, too, the evidence points to topographic features as the pinning
sites. In the Discussion section we revisit the question of why the
pinning energy is so much larger for the aqueous particles than the
oil-dispersed ones. First, however, we examine the nature of the pinning
sites on inorganic particles.

\subsection{Inorganic particles can also pin contact lines}
\begin{figure*}
\centering
  \includegraphics{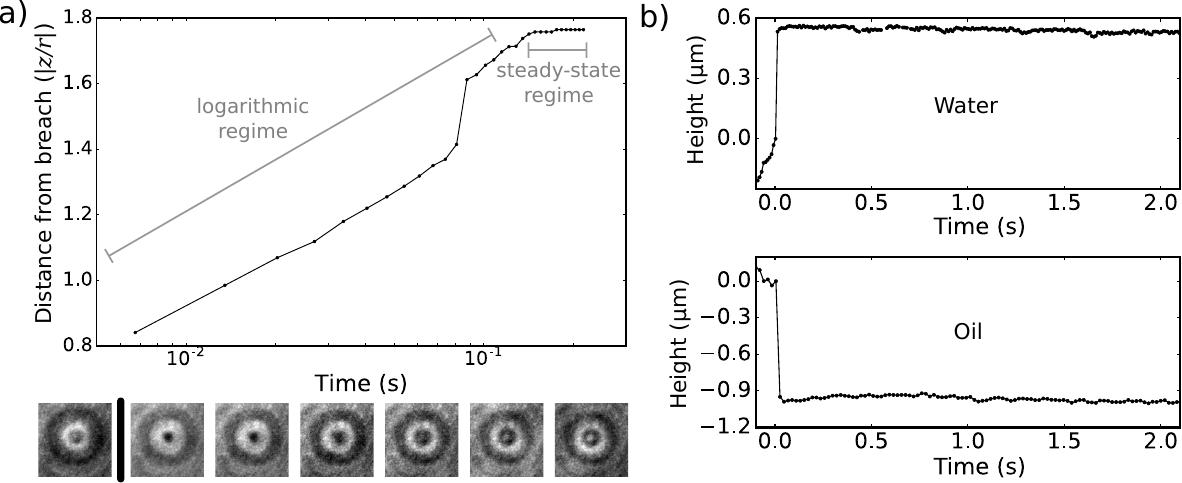}
  \caption[Silica particles breach the interface.]{a) 4-$\mu$m silica
    spheres show logarithmic relaxation after breaching. The insets show
    holograms from 0.0068 s before, then 0.0068 s, 0.0136 s, 0.0271 s,
    0.0542 s, 0.1084 s, and 0.2168 s after the breach. The central
    fringe slowly evolves from dark to bright, indicating a
    wavelength-scale change in the height of the particle. The jump at
    0.8 s is a fast relaxation event, sometimes seen in our samples. b)
    Trajectories of 1-$\mu$m silica spheres approaching the interface
    from the water (top) and decane (bottom) phases. In both cases, the
    particles rapidly reach a steady state height.}
  \label{silica}
\end{figure*}

We find that large, 4 $\mu$m bare silica spheres approaching a
water/glycerol interface from the decane phase relax logarithmically
after breaching but reach a steady-state height after less than 1 s
(Figure~\ref{silica}). These silica particles are large enough that the
slow evolution of the fringe pattern can be detected by eye, as shown in
the insets in Figure~\ref{silica}. Fitting Equation~\eqref{eqnsimple} to
the logarithmic regime yields $A$ = 1 nm$^2$. We calculate $\Delta U$
using Equation~\eqref{crossover} with $z_\mathrm{C}/r$ between 0 and
0.84 and find that the pinning energy is 5--10 $kT$. This value is low
compared to the pinning energies found for aqueous-dispersed polymer
spheres. The result is consistent with the notion that particles that
reach a steady-state contact angles on experimental timescales pin the
contact line with smaller energies.

If we assume the surface asperities are roughly hemispherical caps, we
can compare our fitted $A$ directly with measurements of the roughness
of silica spheres from Ruiz and coworkers,\cite{ruiz_long_2015} who
found the root-mean-squared roughness of 5.2-$\mu$m-diameter silica
particles from Bangs Laboratories to be 1.4 nm using atomic force
microscopy. The size we infer from our dynamic measurements is about 1
nm, in good agreement with the direct measurements.

We also examine 1-$\mu$m bare silica spheres approaching a water-decane
interface from both phases. Most of the particles aggregate when we
attempt to disperse them in decane. To obtain free particles, we discard
the large aggregates that rapidly sediment and dilute the supernatant
with more decane. We do not know whether the surface properties of
silica in water and in decane are the same. However, we find that the
smaller silica spheres reach a steady-state position within 20 ms when
approaching from either phase, as shown in Figure~\ref{silica}. We do
not observe a logarithmic regime, and the spheres reach a steady-state
height within the time resolution of our experiment. Because the
interface in these experiments is not index-matched, the measured height
is only approximate, so we do not calculate a contact angle.

The fast relaxation and absence of any observable logarithmic relaxation
means that we cannot determine if transient pinning or viscous
dissipation sets the rate of relaxation of these spheres. We can,
however, interpret the results in the context of the dynamic models if
we assume that the relaxation is determined by pinning. In that case,
the absence of a logarithmic regime suggests either that the crossover
between the fast and logarithmic relaxation regimes is at timescales
longer than what we can observe or that the difference between
$z_\mathrm{E}$ and $z_\mathrm{C}$ is small. According to
Equation~\eqref{crossover}, a small difference between $z_\mathrm{E}$
and $z_\mathrm{C}$ means that $\Delta U/A$ is small. Indeed,
atomic-force-microscope measurements of similar-sized silica spheres
(0.74 $\mu$m-diameter particles from Duke) by Chen and
coworkers~\cite{chen_attraction_2009} found the RMS roughness to be
about 0.36 nm, which is smaller than the RMS roughness value for larger
silica spheres (1.4 nm).\cite{ruiz_long_2015} Thus one interpretation of
our results is that the small asperities do pin the contact line, but
with a smaller energy than the larger asperities seen on the large
silica spheres, leading to a faster relaxation to equilibrium.

\subsection{Logarithmic relaxation may occur even in sheared emulsion formation}

Finally, we examine whether whether slow relaxation is an important
effect to consider in the preparation of Pickering emulsion, which are
usually made using vigorous mixing. For a 1.9-$\mu$m-diameter
polystyrene sphere with an equilibrium contact angle of 110$^\circ$,
Equation~\eqref{eqn:f} shows that the force on the particle integrated
along the contact line is 10--100 nN for dynamic contact angles between
2$^\circ$ and 107$^\circ$. The force on a 1.9-$\mu$m-diameter particle
in a suspension that is mixed at 11000 rpm in an Ultra Turrax
homogenizer is about 1 nN,\cite{wang_image_2012} orders of magnitude
smaller than the capillary driving force. Thus the relaxation of the
particles is unlikely to be hastened by mixing, and long equilibration
times may be important to take into account in the preparation of
Pickering emulsions. One way to determine the equilibration time is to
vitrify emulsions at different times after formation and image the
interfaces using a method similar to that of Coertjens \textit{et
  al.}~\cite{coertjens_contact_2014}

\section{Discussion}

We have shown that slow relaxation is common to many different kinds of
particles, made of different materials and with different surface
functionalities. Large silica particles and all of our polymer
particles, whether stabilized in water or oil, relax to equilibrium at
rates smaller than those expected from viscous dissipation alone. Thus
we argue that the relaxation rate of colloidal particles at interfaces
is likely controlled by transient pinning and unpinning of the
three-phase contact line.

We have inferred certain features of the pinning sites by fitting
dynamical models that account for pinning and depinning to our data. To
gain further insight into the question of what surface features pin the
contact line we now examine our results across the different types of
systems. Our interpretation assumes that the dynamic models of Kaz
\textit{et al.\@} and Colosqui \textit{et al.\@} capture the essential
physics of the slow relaxation. Although there is little evidence that
viscous dissipation controls the relaxation rate, we cannot---and do not
attempt to---rule out the possibility that more complex wetting
phenomena are responsible for the observed relaxation. Instead we focus
on synthesizing a coherent explanation of the results in the context of
the pinning models.

In all of the systems we observe, the area per defect is inferred to be
on the order of a few square nanometers. This area is comparable to the
area per charged group for aqueous charge-stabilized dispersions, as
noted by Kaz \textit{et al.\@}, but it is much smaller than the expected
area per charged group for non-aqueous, sterically-stabilized polymer
particles such as PHSA-PMMA. In the case of silica spheres, the area per
defect is comparable to the measured surface roughness. We expect
particles with more pronounced surface roughness to be affected more
strongly by contact-line pinning. Taken together, these results suggest
that the pinning sites are small-scale topographical features, perhaps
associated with anchored charged groups in aqueous charge-stabilized
colloids, but not the charges themselves.

In all of the aqueous polymer dispersions, whether charge- or sterically
stabilized, the inferred pinning energy per defect is approximately 50
$kT$. This value contrasts markedly with that of the sterically
stabilized non-aqueous particles and silica spheres, which is only a few
$kT$. To explain this difference we must consider how the surface of the
aqueous polymer spheres differs from that of the non-aqueous polymer
spheres and the silica.

One feature of aqueous polymer spheres that is sometimes mentioned in
the literature is polymer ``hairs''; these are flexible polymer chains
that are attached to the surface of the particles but extend out into
solution and which may contain charged groups. The presence of such
chains was originally inferred from electrophoretic mobility
measurements: Rosen and Saville~\cite{rosen_heatparticles_1992,
  saville_electrokinetic_2000} found that both ``hairy'' polystyrene
particles (with chains grafted onto their surface) and ``bare''
polystyrene particles had much lower electrophoretic mobilities than
those predicted by classical electrokinetic theory. The discrepancy
between experiment and theory was similar for both types of particles,
suggesting that even ``bare'' particles have hairs. For both types of
particles, the agreement between experiment and theory improved
dramatically after the particles were heated past their glass transition
temperature to allow the hairs to anneal to the surfaces of the
particles. Further evidence for polymer hairs comes from optical
measurements of the interaction between a polymer particle and a
surface: Jensenius and Zocchi~\cite{jensenius_measuring_1997} found that
some polystyrene particles attached to surfaces, and, by measuring the
displacement of the particle, they concluded that the attachment tether
was a single polymer chain with a coil size of 50 nm. These experiments
suggest that polymer hairs may be a common feature of polymer particles,
whether there are chains deliberately grafted onto the surface or not.

We therefore hypothesize that polymer hairs are the pinning sites on
aqueous-dispersed polymer particles. Furthermore we hypothesize that the
pinning sites on the non-aqueous polymer particles are also polymer
hairs, which are likely the polymer stabilizers grafted onto the
particles. A possible explanation for why the hairs on the non-aqueous
particles have a much lower pinning energy than the hairs on the aqueous
particles is that the ones on the aqueous particles are
polyelectrolytes. Moving a polyelectrolyte from the aqueous to the oil
phase may involve a large energy barrier because all of the charges need
to first be neutralized. This explanation is not inconsistent with our
results for how the pH affects the relaxation in carboxyl-PS spheres. In
those experiments we found that changing the pH to be closer to the
isoelectric point did not change the area per defect; if the defects are
indeed polyelectrolyte hairs, we expect that some, but not all of the
charges would be neutralized, and so the area per defect (per hair)
would not change. However, the pinning energy should change with the pH.
Therefore this hypothesis can be tested by observing how the crossover
point between exponential and logarithmic relaxation changes as a
function of pH, while independently measuring how the equilibrium
contact angle changes with pH. This is a point for future experiments to
examine. Measurements closer to the isoelectric point could also help to
better isolate the effect of charge on breaching behavior.

\section{Conclusions}

The main message that emerges from our study is that slow, logarithmic
relaxation is a common effect in colloidal particles bound to
interfaces. By ``slow'' we mean slower than the rate expected from
viscous dissipation alone. In many cases, however, the relaxation is
slow even compared to experimental time scales. Our analysis of the
forces involved suggests that the rate of relaxation will not be
significantly altered by vigorous mixing; therefore experiments and
applications (such as making Pickering emulsions) that involve attaching
particles to interfaces and letting them assemble should account for the
possibility that the particles are not in equilibrium on the timescale
of assembly. We expect the out-of-equilibrium behavior to be most
prominent in aqueous polymer particles a few hundred nanometers in
diameter or larger; oil-dispersible polymer particles and silica
spheres, even ones several micrometers in diameter, appear to
equilibrate much more rapidly.

Based on the agreement between the observed logarithmic trajectories and
the predictions of a model based on molecular kinetic theory, we have
argued that the slow relaxation arises from surface heterogeneities that
transiently pin the contact line. We ruled out the possiblity that the
heterogeneities are charged groups directly attached to the surfaces of
the particles. Instead, the likely culprits for the pinning are
topographical features---nanoscale surface roughness in the case of
silica particles and polymer ``hairs'' in the case of polymer particles.
Beyond the implications described above for the assembly of particles at
interfaces, these results also show that the adsorption trajectory is a
sensitive probe of nanoscale surface features that are difficult to
measure directly.

\section{Acknowledgements}
We acknowledge support from the National Science Foundation through
grant number DMR-1306410 and by the Harvard MRSEC through grant number
DMR-1420570. We thank W.B. Rogers for his careful reading of this
manuscript and suggestions; C.E. Colosqui, M. Mani and M.P. Brenner for
critical discussions regarding the interpretation of these results and
the model; A. Hollingsworth for providing the PHSA-stabilized PMMA
particles and for helpful discussions about hairy particles; J.G. Park
for providing the PVP-stabilized PMMA and PVA-stabilized polystyrene
particles, and G. Meng for providing the PDMS-stabilized PMMA particles.
The hologram analysis computations were run on the Odyssey cluster,
supported by the FAS Science Division Research Computing Group at
Harvard University. The zeta potential characterization was performed at
the Center for Nanoscale Systems (CNS), a member of the National
Nanotechnology Coordinated Infrustructure Network (NNCI), which is
supported by the National Science Foundation under NSF award no.
1541959. CNS is part of Harvard University.


\bibliography{rsc} 
\bibliographystyle{rsc} 

\end{document}